\documentclass{article}
\usepackage{spconf,amsmath,graphicx}
\usepackage{multirow, tabularx}
\usepackage{booktabs}
\usepackage{xcolor}
\usepackage{amssymb}
\usepackage{enumitem}
\setitemize{noitemsep,topsep=0pt,parsep=0pt,partopsep=0pt}
\usepackage{algorithm}
\usepackage{algpseudocode}

\title{Deep AHS: A Deep Learning Approach to Acoustic Howling Suppression}
\name{Hao Zhang, Meng Yu, Dong Yu}
\address{Tencent AI Lab, Bellevue, WA, USA}

\begin{document}
\ninept
\maketitle
\begin{abstract}
In this paper, we formulate acoustic howling suppression (AHS) as a supervised learning problem and propose a deep learning approach, called Deep AHS, to address it. 
Deep AHS is trained in a teacher forcing way which converts the recurrent howling suppression process into an instantaneous speech separation process to simplify the problem and accelerate the model training. 
The proposed method utilizes properly designed features and trains an attention based recurrent neural network (RNN) to extract the target signal from the microphone recording, thus attenuating the playback signal that may lead to howling. 
Different training strategies are investigated and a streaming inference method implemented in a recurrent mode used to evaluate the performance of the proposed method for real-time howling suppression.
Deep AHS avoids howling detection and intrinsically prohibits howling from happening, allowing for more flexibility in the design of audio systems.
Experimental results show the effectiveness of the proposed method for howling suppression under different scenarios.


\end{abstract}
\section{Introduction}
\label{sec:intro}


Howling arises due to the coupling between the microphone and loudspeaker when there exists positive feedback \cite{waterhouse1965theory, van2010fifty}. Specifically, the microphone signal in an audio system is played out through a loudspeaker that is exposed in the same space and then picked up again by the same microphone, forming a closed acoustic loop. If not properly handled, this playback signal will be looped back repeatedly and result in shrill sound at frequencies that have unity or larger loop gain. This phenomenon is known as howling. 
Howling is a crucial problem for video/audio conferences and acoustic amplification systems such as hearing aids and karaoke. It is not only harmful to our auditory system but also destructive to the amplification equipment \cite{agnew1996acoustic}. 

Many acoustic howling suppression (AHS) solutions have been proposed to address this problem, including gain control \cite{schroeder1964improvement, berdahl2010frequency}, notch filter (NF) \cite{loetwassana2007adaptive, gil2009regularized, waterschoot2010comparative}, and adaptive feedback cancellation (AFC) \cite{joson1993adaptive}.
The gain reduction method can be achieved by either manually reducing the volume of an amplifier or altering the position of audio devices. However, such methods are with restricted applications and unsuitable in scenarios that require high acoustic amplification \cite{lee2011improvements}.
The NF methods attenuate howling by adjusting their filter coefficients to form a null at frequencies where howling appears. 
However, the NF methods require accurate detection of howling and inherently distort the target sound and even introduce unexpected howling frequencies \cite{loetwassana2007adaptive}.
AFC attenuates howling by estimating the acoustic path between the loudspeaker and microphone using adaptive filters. Because the target signal and playback signal are highly correlated, de-correlation techniques are usually required in AFC methods,
which, however, usually distorts speech quality \cite{wang2020adaptive}. 

%
%
%
%

The ultimate goal of howling suppression is to attenuate the playback signal and send only the target signal to the loudspeaker, which is essentially the same as how we address acoustic echo cancellation (AEC). 
Considering that deep learning is powerful at modeling complex nonlinear relationships and has been successfully introduced to suppress acoustic echo \cite{zhang2018deep, zhang2019deep, zhang2021ft, cutler2022icassp,  zhang2022neural}, it could also serve as a powerful alternative to address AHS problem. Chen et al. \cite{chen2022neural} proposed a deep learning based method for howling detection. Gan et al. \cite{gan2022howling} employed deep learning for howling noise suppression. However, the method presented in \cite{gan2022howling} treats howling as a type of noise for speech enhancement rather than suppressing howling in a streaming and recurrent manner.


In this paper, we propose a deep learning approach, called Deep AHS, to address howling suppression. Our approach formulates AHS as a supervised learning problem and the overall task is to maintain only the target signal while suppressing the playback signal and background noise in the microphone recording. 
Considering that playback signal and target signal are highly correlated, we use a concatenation of temporal correlation, frequency correlation, and channel covariance of input signals as feature and train an attention based recurrent neural network (RNN) \cite{yu2022neuralecho} to estimate a complex ratio filter \cite{mack2019deep} of the target signal. To the best of our knowledge, this is the first study that employs deep learning for acoustic howling suppression.
The contribution of this study is fourfold. 
Firstly, Deep AHS formulates howling suppression, an adaptive procedure, as a supervised learning problem with the help of teacher-forced learning. It is fundamentally different from traditional AHS methods and does not require howling detection.
Secondly, with such a training strategy, a streaming inference method is implemented to evaluate the performance of Deep AHS in a recurrent manner.
Thirdly, Deep AHS is robust to nonlinear distortions and can achieve howling and noise suppression jointly under different scenarios, which allows for higher loop gain and brings flexibility to the design of an audio system.
Lastly, different training strategies are investigated and systematically compared in this study to show the efficiency of our method for howling suppression.


The remainder of this paper is organized as follows. Section 2 introduces acoustic howling. Section 3 describes our proposed method. The experimental setup and results are introduced in Section 4 and Section 5, respectively. Section 6 concludes the paper.

\section{Formulation of Acoustic Howling}
\label{AHS}

\begin{figure}[!t]
\centering
     \includegraphics[width=0.9\columnwidth]{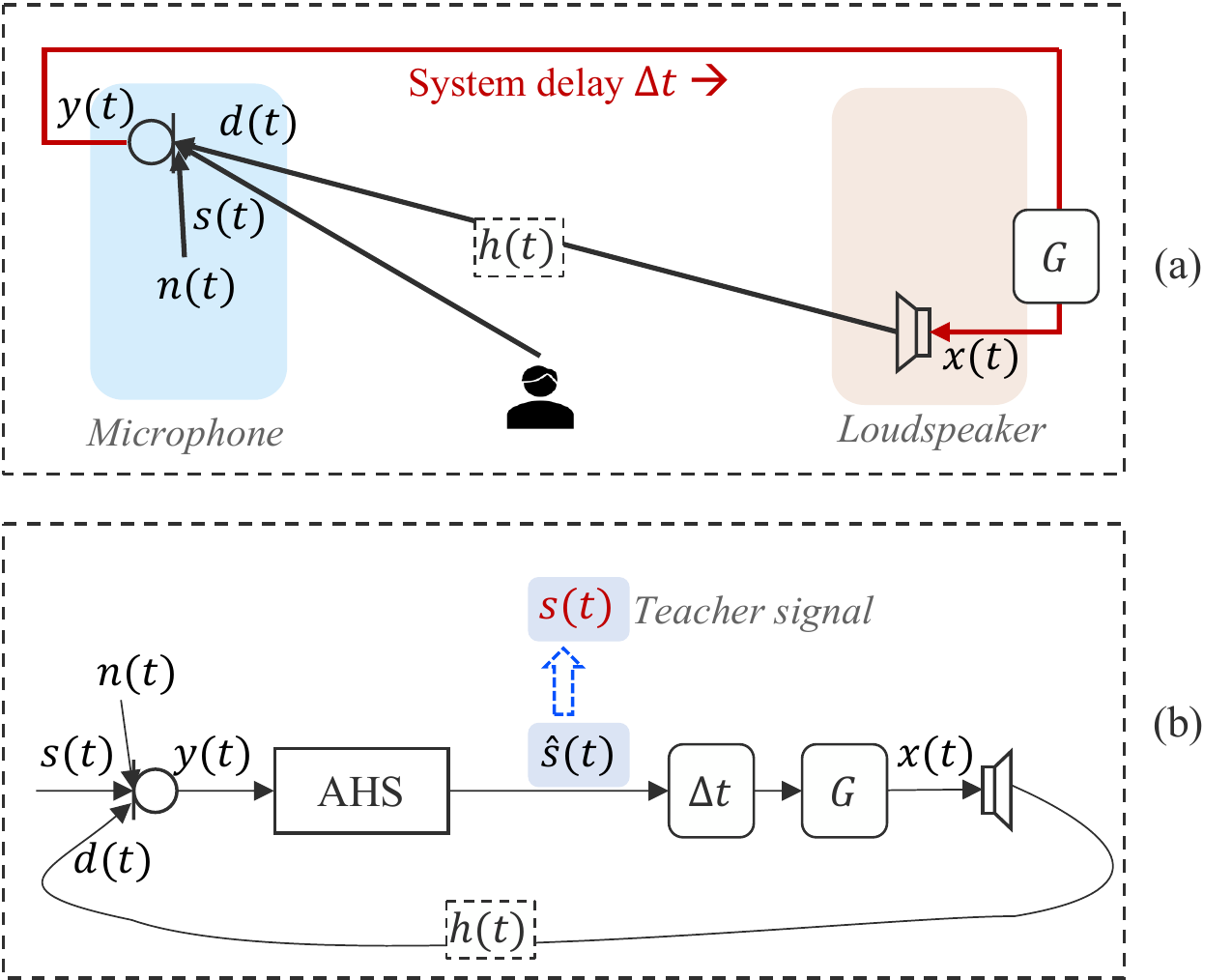}
      \caption{Diagrams of (a) an acoustic amplification system, (b) signal flow in an acoustic howling suppression system.}
      \label{fig:AcousticModels}
\end{figure}

Without loss of generality, let us consider a typical single-channel acoustic amplification system with a microphone and a loudspeaker coupled in the same space, as shown in Fig.~\ref{fig:AcousticModels}(a). 
The target speech is picked up by the microphone as $s(t)$, which is then sent to the loudspeaker for acoustic amplification.
The loudspeaker signal $x(t)$ is played out and arrives at the microphone as a playback signal denoted as $d(t)$:
\begin{flalign}
d(t) = \textstyle NL[x(t)]*h(t)
\end{flalign}
where $NL(\cdot)$ denotes the nonlinear distortion introduced by the loudspeaker, $h(t)$ represents the acoustic path from loudspeaker to microphone, and $*$ denotes linear convolution.

If without any processing, the loudspeaker signal $x(t)$ will be a delayed and amplified version of $y(t)$, and this playback signal $d(t)$ will re-enter the pickup repeatedly, the corresponding microphone signal at time index $t$ can be represented as:
\begin{flalign}
\label{equ:howling}
\textstyle y(t) =   s(t) + n(t) +  NL\left[y(t-\Delta t) \cdot G\right]*h(t) 
\end{flalign}
where $n(t)$ represents the background noise, $\Delta t$ denotes the system delay from microphone to loudspeaker, and $G$ the gain of amplifier.
The recursive relationship between $y(t)$ and $y(t-\Delta t)$ causes re-amplifying of playback signal and will form positive feedback and result in circular amplification at certain frequencies, known as acoustic howling.
With that being said, howling is generated in a recurrent manner rather than instantaneously. It starts as multiple playback signals and gradually forms a shrill sound after being amplified to a certain extent. 

It is worth noting that acoustic howling is different from acoustic echo even though inappropriately handled acoustic echo (leakage) could also result in howling. The major differences between them are:
1. Both of them are essentially playback signals while howling is generated gradually. 
2. The playback signal that leads to howling is generated from the same source as that of the target signal. While acoustic echo is usually generated from a different source (far-end speaker), which makes the suppression of howling more challenging.


%
%
%

%

\section{Proposed Method: Deep AHS}

\subsection{Teacher-forced learning for howling suppression}


%
%
Ideally, if the AHS method can always perfectly process microphone recording and completely attenuates the playback component in it before sending it to the loudspeaker, there will be no howling problem under any circumstances. 
From the speech separation point of view, it seems that AHS can be seen as a speech separation problem where the target signal $s(t)$ is a source to be separated from the microphone signal, which is similar to the idea of how deep learning based AEC is formulated.

However, it is non-trivial to achieve howling suppression using deep learning considering the characteristics of acoustic howling described in Sec.~\ref{AHS}. The most crucial problem is that howling is generated adaptively, and the current input depends on the previous outputs.
Specifically, the existence of distortion/leakage in the current processed signal $\hat{s}(t)$, as shown in Fig.~\ref{fig:AcousticModels}(b), will affect the playback signal received at the microphone in the next loop $d(t+\Delta t)$.
Ideally, we should train a deep learning model in an adaptive way by updating its parameters on a sample level. However, this requires a huge amount of computation and is hard to be realized in real applications. 

We propose Deep AHS to train a model for howling suppression using teacher-forced learning \cite{williams1989learning, lamb2016professor}.
Assuming that once the model is properly trained, it should attenuate the playback signal in the microphone and send only target speech to the loudspeaker. 
During model training, we take the target speech, $s(t)$, as the teacher signal to replace the actual output $\hat{s}(t)$ in the subsequent computation of the network, as shown in Fig.~\ref{fig:AcousticModels}(b). 

By using teacher forcing learning, the playback signal $d(t)$ is then a determined signal influenced only by $s(t)$, and the repeating summation of multiple playback signals in Eq. (\ref{equ:howling}) can be simplified to a one-time playback. 
The corresponding microphone signal for model training can be written as:
\begin{flalign}
\label{equ:mic}
y(t)  \textstyle&= s(t) + n(t) + NL[s(t-\Delta t) \cdot G] *h(t)
\end{flalign}
The microphone signal during teacher forcing learning is a mixture of the target signal, background noise, and a determined one-time playback signal. And the overall problem can thus be formulated as a speech separation problem.
Training Deep AHS in a teacher-forced learning way not only simplifies the overall problem but also possible to diminish the uncertainty introduced in the adaptive process of AHS and results in a robust howling suppression solution.


\subsection{Different training strategies}

Different training strategies have been explored in this paper. 
The most straightforward one is to directly use the microphone signal in (\ref{equ:mic}) as input and set the corresponding $s(t)$ as the training target. We name this training strategy as the model trained without using a reference signal (``w/o Ref''). 

A more proper way is to extract more information from input and use it as a reference signal during model training. 
We propose to use a delayed microphone signal as additional input (reference signal) with the amount of delay estimated during an initial stage. 
Considering that the playback signal can be regarded as a delayed, scaled, nonlinear version of $s(t)$, using a delayed microphone signal helps the model to better differentiate the target signal from playback. We name this training strategy as ``w Ref''.

In addition, admitted that there is always a mismatch during offline training and real-time application considering the leakage existed in $\hat{s}$.
To incorporate the mismatch and better approximate the real scenarios, we propose another strategy that works by fine-tuning the model using pre-processing signals, denoted as ``Fine-tuned''.
Then, the microphone signal for offline training is a modified version of (\ref{equ:mic}):
\begin{flalign}
 \textstyle y'(t) = s(t) + d'(t) + n(t)
\end{flalign}
where $d'(t)$ is the distorted playback signal generated using estimated target $\hat{s}(t-\Delta t)$. To be specific, we pre-process all the training data using a pre-trained model and then feed the enhanced output through the audio system to get the corresponding playback $d'(t)$. Finally, we fine-tune the model using $y'(t)$ as input. 
It is expected that the mismatch mentioned previously would be reduced slightly given that the model has seen the distortion during training. 

%
%

%

\subsection{Details of network structure}
\begin{figure}[!t]
\centering
     \includegraphics[width=0.9\columnwidth]{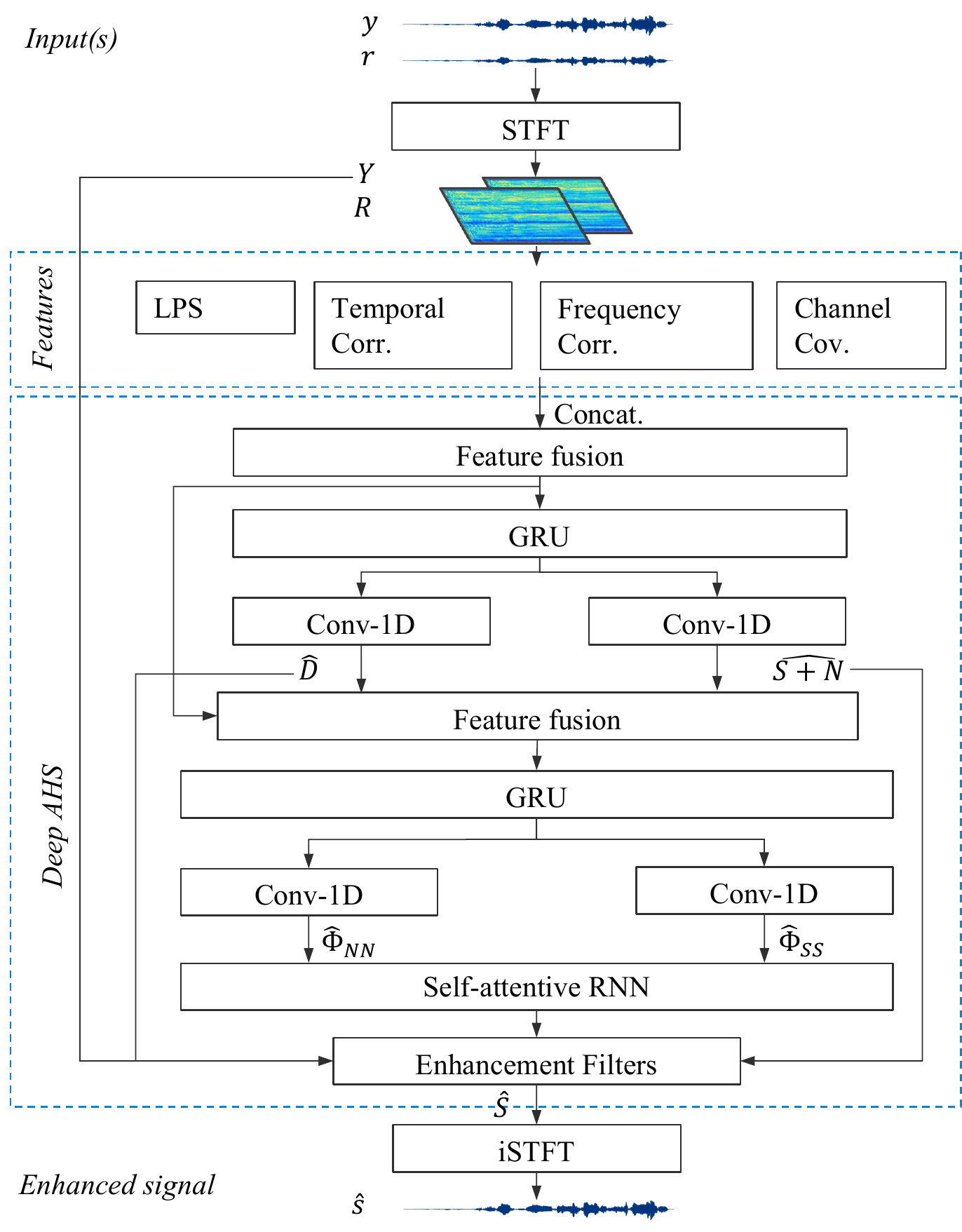}
      \caption{The architecture of Deep AHS for howling suppression.}
      \label{fig:NeuralEcho}
\end{figure}

A diagram of the proposed method is shown in Fig.~\ref{fig:NeuralEcho}
The microphone signal $y(t)$ and reference signal $r(t)$, sampled at 16k Hz, are firstly divided into 32-ms frames with 16 ms frameshift. A 512-point short-time Fourier transform (STFT) is then applied to each frame, resulting in the frequency domain inputs, $Y$ and $R$.
Besides the normalized log-power spectra (LPS), we calculate the correlation matrix across time frames and frequency bins of microphone and reference signals, respectively, as input features. 
These features are used to capture the temporal and frequency characteristics of input signals and are helpful to discriminate between howling and tonal components. 
Channel covariance of input signals is calculated as another input feature to account for cross-correlation between them. A concatenation of these features is used for model training with a linear layer for feature fusion.


The network consists of three parts, where the first part employs a gated recurrent unit (GRU) layer with 257 hidden units and two 1D convolution layers to estimate a complex-valued filter for playback suppression and playback estimation, respectively. The estimates are then applied on the input signals through deep filtering \cite{mack2019deep} to obtain the corresponding intermediate outputs, denoted as $\tilde{Y}$ and $\tilde{R}$. 

These intermediate outputs are then used as additional inputs and their LPS, together with the fused feature for the first part, are concatenated and fused to serve as the inputs for for another GRU layer. 
We regard $Y$, $\tilde{Y}$, and $\tilde{R}$ as three-channel inputs and employ two 1D convolution layers for each input channel to estimate the playback/noise and target speech components in it. The corresponding covariance matrices of playback/noise $\hat{\Phi}_{NN}$ and target speech $\hat{\Phi}_{SS}$ are calculated and concatenated as the input to the third part. 

The third part is for enhancement filter estimation, which is motivated by the idea of multi-channel signal processing. We train a self-attentive RNN to estimate a three-channel speech enhancement filters $\mathbf{W}\in \mathbb{C}^{F\times T \times 3}$. These filters are then applied on the input channels ($Y$, $\tilde{Y}$, and $\tilde{R}$) to get the enhanced target speech $\hat{S}$. Finally an inverse STFT (iSTFT) is used to get waveform $\hat{s}$. Details of the feature design and network structure can be found in \cite{yu2022neuralecho}.

The loss function for model training is defined as a combination of scale-invariance signal-to-distortion ratio (SI-SDR) \cite{le2019sdr} in the time domain and mean absolute error (MAE) of spectrum magnitude in the frequency domain:
\begin{flalign}
Loss = -\text{SI-SDR}(\hat{s}, s) + \lambda \text{MAE} (|\hat{S}|, |S|)
\end{flalign}
where $\lambda$ is set to 10000 to balance the value range of the two losses.

\subsection{Streaming inference in a recurrent mode}

\begin{algorithm}
\caption{Streaming inference process.}\label{alg:streaming}
\begin{algorithmic}
  \Procedure{Streaming}{$(y, r) \rightarrow \hat{s}$}  \Comment{$r$: reference signal}

    \State Initialization: $\mathbf{y}_\texttt{buffer} $, $\mathbf{r}_\texttt{buffer}$, load DNN model: $f(\cdot)$ 
    \State Parameters: Delay: $\Delta t$, gain: $G$, nonlinearity: $NL(\cdot)$
    \While{$m\leq M$} \Comment{$M$ is the total number of frames}
      \State $\hat{s}_m \gets f(\mathbf{y}_\texttt{buffer}, \mathbf{r}_\texttt{buffer})$  \Comment{Get output at frame $m$}
      \State  $\hat{s} \gets \hat{s}_m$  \Comment{Save to final output}
      \State $x_{m+1} \gets \text{Delayed}\_\hat{s}_{m}\cdot G$  \Comment{Get loudspeaker signal}
      \State $d_{m+1} \gets NL(x_{m+1})*h$  \Comment{Get next playback}
      \State $\mathbf{y}_\texttt{buffer} \gets d_{m+1} + s_{m+1} + n_{m+1}$ \Comment{Update input}
      \State  $\mathbf{r}_\texttt{buffer} \gets \mathbf{y}_\texttt{buffer}$ \Comment{Update reference signal}
    \EndWhile\label{euclidendwhile}
    \State \textbf{return} $\hat{s}$\Comment{The streaming output is $\hat{s}$}
  \EndProcedure
  \end{algorithmic}

\end{algorithm}

Since there is always a mismatch between the offline training and inference stage of Deep AHS. 
A streaming inference method, in which the output of the processor is looped back and added to the input in the following time steps, is therefore implemented to evaluate the performance of Deep AHS in a realistic and recurrent mode.
Details of this streaming inference are shown in Algorithm 1.

%
%
%
%

\section{Experimental Setup}

\subsection{Data preparation}

%
%

We simulate the single-channel dataset with playback and background noise using
AISHELL-2 \cite{du2018aishell}. 
A total number of 10k room impulse responses (RIRs) are generated using the image method \cite{allen1979image} with random room characteristics and reverberation time (RT60) ranging from 0s to 0.6s. Each RIR is a
set consisting of RIRs for the near-end speaker, loudspeaker and
background noise locations. 
During data generation, a randomly selected RIR set is utilized to generate target speech $s$ and its corresponding one-time playback signal $d$ using system delay $\Delta t$ randomly generated within the range of $[0.1, 0.5]$ seconds. The nonlinear distortions introduced by the amplifier and loudspeaker are simulated as a saturation type of nonlinearity using hard clipping and Sigmoidal function \cite{zhang2018deep}. The microphone signal for offline training is created as a mixture with signal-to-playback-ratio (SPR) randomly selected from $[-15, 20]$ dB and signal-to-noise ratio (SNR) ranging from $-10$ dB to $40$ dB.
A total number of 10k, 0.3k, and 0.5k utterances are generated for training, validation and testing, respectively. The testing data are generated using different utterances and RIRs from that of the training and validation data.  The model is trained for 30 epochs with a batch size of 20. 

\subsection{Method evaluation}
%
%

The performance of the proposed method is evaluated in two different manners: offline evaluation and streaming inference.
The offline evaluation uses signals generated in (\ref{equ:mic}) as input to evaluate playback attenuation performance. SI-SDR and perceptual evaluation of speech quality (PESQ) \cite{rix2001perceptual} are used to evaluate the extent of playback attenuation and quality of target speech. And a higher value denotes better performance. 

For streaming inference, we generate the enhanced signal using the streaming method described in Algorithm 1, where the microphone signal is updated in a recurrent mode with the leakage/distortion taken into consideration. This inference method mimics the manner of howling generation in real applications and can be used to show real-time howling suppression performance. 

\section{Experimental Results}

\subsection{Offline evaluation for playback attenuation}

%

%


\begin{table}[t]
\centering
\caption{Offline evaluation of models for playback attenuation.}
\label{table:offline}
\resizebox{0.45 \textwidth}{!}{
\begin{tabular}{l|ccc|ccc} \specialrule{1.5 pt}{1 pt}{1 pt}
Models            & \multicolumn{3}{c|}{SI-SDR} & \multicolumn{3}{c}{PESQ} \\ \hline
SPR               & -5      & 0       & 5      & -5     & 0      & 5      \\ \hline
unprocessed: $y$               & -4.29   & 0.72    & 5.70   & 1.66   & 2.06   & 2.45   \\
w/o Ref           & 6.78    & 9.88    & 13.02  & 2.61   & 2.90   & 3.17   \\
w Ref             & 9.19    & 12.12   & 15.22  & 2.81   & 3.09   & 3.33   \\   \hline \hline

unprocessed: $y'$ & -3.01   &  1.47  &  5.98 & 1.83   &  2.24  &  2.58   \\
w Ref, fine-tuned  & 10.10   & 13.96  &  18.06 & 3.05 &  3.37  &  3.60  \\  \specialrule{1.5 pt}{1 pt}{1 pt}
\end{tabular}
}
\end{table}



We start by evaluating the proposed method in an offline manner. To be specific, the microphone signals used for testing are generated in the same way as those used during model training, the results tested under different SPR levels and 30 dB SNR are shown in Table~\ref{table:offline}. 
It is observed that the proposed method with only a microphone signal as input attenuates the playback signal efficiently. Using a delayed microphone signal as a reference signal helps further improve the overall performance.
It is worth noting that the results shown in Table~\ref{table:offline} only demonstrate that the proposed method is capable of extracting target speech and attenuating the one-time playback in the microphone recording.
In Sec.~\ref{results:2}, we will present the results obtained using streaming inference to show its performance for howling suppression during real-time implementation.

In addition, the results obtained using the fine-tuned model are presented in the last two rows Table~\ref{table:offline}. 
Since the fine-tuned model is tested using pre-processed signals, which makes it improper to directly compare the fine-tuned model with the other two in this evaluation manner. 
However, it is fair to make such a comparison during streaming inference since the model with and without fine-tuning are handled in the same way during streaming inference. There is no need to do pre-processing for the fine-tuned model.


\subsection{Streaming inference for howling suppression}
\label{results:2}
%
%
%


\begin{table}[]
\centering
\caption{PESQ of streaming evaluation for howling suppression.}
\label{table:streaming}
\resizebox{0.49 \textwidth}{!}{
\begin{tabular}{l|cccccc} \specialrule{1.5 pt}{1 pt}{1 pt}
 Playback level  & no AHS & NF  &AFC & w/o Ref & w Ref & w Ref, Fine-tuned  \\ \hline
Soft   &  2.45   &  2.50 & 2.60  &   2.74     &  2.95      &     3.02                 \\ 
Moderate   &   1.78  & 1.85 & 2.22   &   2.27     &   2.68     &    2.74                  \\  
Severe   &  1.63  & 1.79 &  2.13   &   2.07    &  2.53  &    2.56               \\ \specialrule{1.5 pt}{1 pt}{1 pt}
\end{tabular}
}
\end{table}



\begin{figure}[!t]
\centering
     \includegraphics[width=0.99\columnwidth]{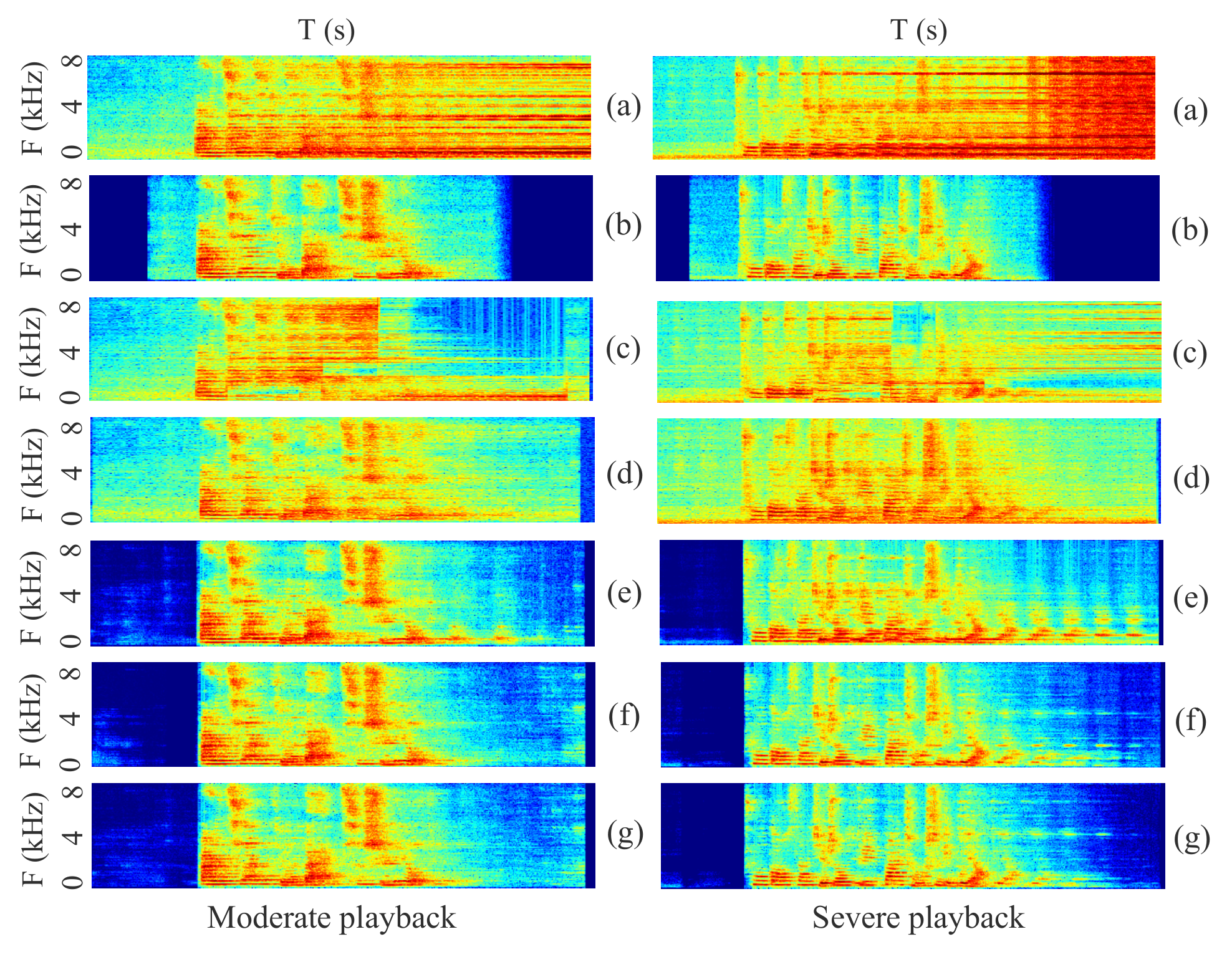}
      \caption{Spectrograms of different methods for howling suppression with moderate and severe howling. In each panel: (a) no AHS, (b) target signal, (c) NF, (d) AFC, (e) w/o Ref, (f) w Ref, (g) w Ref, fine-tuned.}
      \label{fig:spectrograms}
\end{figure}

This part evaluates the performance of Deep AHS for real-time howling suppression and the outputs are generated using the streaming inference method described in Algorithm 1. 
We generate three testing scenarios with soft, moderate, and severe playback by gradually increasing the amplification gain $G$ during the streaming stage and present the results in Table~\ref{table:streaming}. 
To better show the howling suppression performance, spectrograms of two testing samples are shown in Fig.~\ref{fig:spectrograms}.
Notch filter \cite{loetwassana2007adaptive} and adaptive feedback cancellation based AHS using Kalman filter \cite{albu2018hybrid} are utilized for comparison.
It is observed that the proposed method is capable of suppressing howling and Deep AHS models with reference signal consistently outperform traditional AHS methods.
The fine-tuned model achieves the best overall performance. 

\section{Conclusion}

In this paper, we have proposed, for the first time, a deep learning approach to acoustic howling suppression. The proposed method addresses AHS by extracting the target signal from microphone recording using an attention based recurrent neural network with properly designed features. With the idea of teacher-forced learning, the Deep AHS model is trained offline using teacher signals and evaluated in both offline and streaming manners to show its performance for howling suppression.
Multiple training strategies have been investigated and the evaluation results under different scenarios show the effectiveness of the proposed method for howling suppression.
Future work includes device implementation of Deep AHS, considering acoustic scenarios with both acoustic howling and acoustic echo, and expanding the proposed method for multi-channel systems.

%
%
%

\vfill\pagebreak


\bibliographystyle{IEEEbib}
\bibliography{HowlingBIB}

\end{document}